# Bridging magnonics and spin-orbitronics

Boris Divinskiy[1*], Vladislav E. Demidov[1], Sergei Urazhdin[2], Ryan Freeman[2], Anatoly B. Rinkevich[3] & Sergej O. Demokritov[1,3]

[1]*Institute for Applied Physics and Center for Nonlinear Science, University of Muenster, 48149 Muenster, Germany*

[2]*Department of Physics, Emory University, Atlanta, GA 30322, USA*

[3]*Institute of Metal Physics, Ural Division of RAS, Ekaterinburg 620990, Russian Federation*

**The emerging field of nano-magnonics utilizes high-frequency waves of magnetization – the spin waves – for the transmission and processing of information on the nanoscale. The advent of spin-transfer torque has spurred significant advances in nano-magnonics, by enabling highly efficient local spin-wave generation in magnonic nanodevices. Furthermore, the recent emergence of spin-orbitronics, which utilizes spin-orbit interaction as the source of spin torque, has provided a unique ability to exert spin torque over spatially extended areas of magnonic structures, enabling enhanced spin-wave transmission. Here, we experimentally demonstrate that these advances can be efficiently combined. We utilize the same spin-orbit torque mechanism for the generation of propagating spin waves, and for the long-range enhancement of their propagation, in a single integrated nano-magnonic device. The demonstrated system exhibits a controllable directional asymmetry of spin wave emission, which is highly beneficial for applications in non-reciprocal magnonic logic and neuromorphic computing.**

The advancement of nanomagnonic circuits requires the ability to downscale the individual circuit elements, which can be enabled by the new methods of local spin wave generation and guiding in nanostructures[1-4]. Meanwhile, increasing complexity of integrated magnonic circuits requires enhanced coherence and propagation length of spin waves. Both of these challenges can be addressed by utilizing spin torque[5,6], which provides the ability to electrically control magnetic damping, resulting in enhanced spin wave propagation, and to completely compensate damping, resulting in the local generation of coherent spin waves.

The possibility to locally generate spin waves by spin torque has been demonstrated for devices operated by spin-polarized electric currents[7-10], and by pure spin currents produced by nonlocal spin injection[11]. All these devices utilized current injection through a magnetic nano-contact, resulting only in local spin torque that could not provide long-range enhancement of spin wave propagation.

In contrast to the conventional spin torques produced by the local current injection through conducting magnetic materials, the spin-orbit torque (SOT)[12,13] provides the ability to compensate magnetic damping in spatially extended regions of both conducting and insulating magnetic materials[14-16]. This advantage is particularly significant for magnonics, since it enables long-range enhancement of spin-wave propagation in a variety of materials suitable for the implementation of magnonic nano-devices[17-21].

While SOT-induced coherent magnetic auto-oscillation has been achieved in several nanomagnetic device geometries[22,23], and many other novel SOT oscillators have been proposed[16,24-27], none of them provided the possibility to generate coherent propagating spin waves. In these systems, the SOT-induced coherent dynamics was spatially confined due to the nonlinear self-localization phenomena and/or the inhomogeneity of the effective magnetic field. Meanwhile, SOT-enhanced spin-wave propagation was demonstrated for structures that are not suitable for spin wave generation[17-21]. However, simultaneous local generation of spin waves and long-range enhancement of their propagation has not been achieved so far.

Here, we experimentally demonstrate a magnonic nano-device utilizing the same SOT mechanisms to generate propagating spin waves, and to simultaneously compensate their propagation losses over a spatially extended region. The system is based on a new concept of nano-notch spin-Hall auto-oscillator directly incorporated into a magnonic nano-waveguide. The oscillator is capable of unidirectional emission of spin waves, controlled by the direction of the static magnetic field. The propagation length of spin-waves emitted into the waveguide is enhanced by SOT by up to a factor of three, which can be further increased by the material engineering and geometry optimization. The proposed approach can be easily scaled to chains of SOT nano-oscillators coupled via propagating spin waves, facilitating the development of novel nanoscale signal processing circuits such as logic and neuromorphic computing networks[28].

Our test devices (Fig. 1) are based on 180 nm-wide and 4 µm-long nano-waveguides patterned from a Permalloy(Py)(15 nm)/Pt(4 nm) bilayer. Ion milling was used to pattern a rectangular 200 nm-wide and 10 nm-deep notch in the top Py layer of the waveguide, forming a nano-oscillator that serves as the spin-wave source. When electric current $I$ flows through the device, the spin-Hall effect (SHE)[29,30] in Pt injects pure spin current $I_S$ into the Py layer, producing spin-orbit torque on its magnetization $M$ that compensates the magnetic damping[14]. To maximize the effect of SOT, the static magnetic field $H_0$ is applied perpendicular to the direction of the flow of the electric current $I$, in accordance with the symmetry of SHE. The thickness-averaged magnitude of the anti-damping torque is inversely proportional to the thickness of the magnetic layer[5,6]. Thus, the effects of spin torque on the 5 nm-thick Py layer in the nano-notch area are significantly larger than on the 15 nm-thick Py waveguide. As the current $I$ is increased, damping becomes completely compensated in the nano-notch region, resulting in the local excitation of magnetization auto-oscillations. Meanwhile, damping remains only partially compensated in the waveguide. The nano-notch oscillator can emit spin waves into the waveguide, provided that the latter supports propagating spin-waves at the frequency of auto-oscillation. Since the entire waveguide is subjected to the spin current $I_S$ reducing the natural damping, the propagation of the emitted spin waves can become enhanced by the current-induced SOT.

We study the SOT-induced magnetization dynamics by using the micro-focus Brillouin light scattering (BLS) spectroscopy[31]. We focus the probing laser light into a diffraction-limited spot on the surface of the studied structure (Fig. 1). The intensity of light inelastically scattered by the magnetic oscillations – the BLS intensity – is proportional to the intensity of magnetic oscillations at the position of the focal spot, while its frequency is determined by that of the oscillations.

First, we characterize the auto-oscillation by analyzing the BLS spectra obtained with the laser spot positioned on the nano-notch region. A representative spectrum obtained at $I$=3.8 mA exhibits two intense auto-oscillation peaks (Fig. 2a). Micromagnetic simulations described below allow us to identify these peaks as the two fundamental dynamic modes of the nano-notch, characterized by different distributions of the dynamic magnetization across the nanowire width (insets in Fig. 2a). For the low-frequency (LF) mode, the dynamical magnetization amplitude is largest at the edges of the nanowire, while for the high-frequency (HF) mode, the amplitude is largest at the center.

The device exhibits stable auto-oscillations over a significant range of driving currents (Fig. 2b). The characteristic current density in Pt of $2\text{-}3\times10^{12}$ A/m$^2$ is comparable to other SOT-based nano-oscillators[16]. The lower-frequency LF mode starts to auto-oscillate at smaller currents than the HF mode due to its lower relaxation rate, in agreement with the theory of spin torque auto-oscillators[32]. Both modes exhibit a red nonlinear frequency shift with increasing $I$, consistent with the effects of spin current on the dynamical spectrum. Both modes are characterized by a "soft" auto-oscillation onset - a gradual initial increase of intensity with increasing driving current (Fig. 2c). The intensity of the LF mode saturates and starts to decrease around $I$=3.6 mA, while the intensity of the HF mode exhibits a rapid increase. These behaviors can be attributed to the competition between the modes for the angular momentum supplied by the spin current.

To analyze the oscillation-induced magnetization dynamics in the waveguide, we raster the probing BLS spot over the waveguide area. Figures 3a and 3b show representative maps of the dynamic magnetization at the LF and the HF mode frequency, respectively. These data demonstrate that the LF mode is localized in the nano-notch, and does not emit spin waves into the waveguide. In contrast, the HF mode emits spin waves, preferentially in the negative-$x$ direction. To characterize the emission and propagation quantitatively, in Fig. 3c we plot on the log-linear scale the propagation-coordinate dependence of the BLS intensity. These data clearly demonstrate that the spin waves emitted by the HF mode exponentially decay away from the nano-notch. By fitting the data with the exponential function, we find the decay length $L$=1.5 μm, at which the amplitude of the wave decreases by a factor of e. By comparing the intensities of the waves emitted to the left and to the right, we also determine that the decay length is the same for the two directions, but the intensities differ by about a factor of 3.

Additional measurements show that the direction of the preferential emission can be reversed by reversing the direction of the static magnetic field (Fig. 3d). To achieve auto-oscillations, the direction of the dc current is also reversed (see arrows in Figs. 3b and 3d), in accordance with the symmetry of SHE. Thus, the demonstrated structure provides the ability to control unidirectional emission of spin waves by the magnetic field.

We use micromagnetic simulations to gain insight into the mechanisms of directional spin wave emission and propagation. The oscillation of magnetization in the nano-notch area is simulated with a local monochromatic microwave field, applied perpendicular to the surface. Figure 4a shows snapshots of the dynamic magnetization in the waveguide produced by the excitation at the frequencies of the LF and the HF modes, as labeled. Oscillations excited at the frequency of the LF mode remain localized in the notch area and do not generate propagating spin waves, in agreement with the experimental data. This result indicates that the waveguide does not support spin-wave propagation at this frequency. In contrast, excitation at the frequency of the HF mode results in the generation of spin waves propagating in both directions away from the notch.

Figure 4b shows the current-dependent values of the spin-wave decay length obtained in the experiment and from the simulations. For each value of $I$, the excitation frequency used in the simulations is determined from the experimental data (Fig. 2b), and the damping constant is set to the standard value for Py, $\alpha_0=0.01$. The simulations reproduce the experimentally observed reduction of the decay length with increasing current, which can be attributed to the smaller group velocity of lower-frequency spin waves excited at larger currents[31]. However, the magnitude of the decay length observed in the experiment is significantly larger than that obtained in the simulations using the natural damping constant. The ratio of the two lengths increases from 2 at $I=3.6$ mA to 3 at $I=4.5$ mA (symbols in Fig. 4c).

We attribute the observed difference to the compensation of the magnetic damping in the waveguide by the spin current injected over the entire spin-wave propagation path (Fig. 1). According to the spin torque theory[32], the current-dependent effective damping varies as $\alpha(I) = \alpha_0(1 - I/I_C)$, where $\alpha_0$ is the damping constant at $I=0$, and $I_C$ is the critical current, at which the damping becomes completely compensated. By using the spectroscopy of thermal magnetic fluctuations (see Supplementary Information), we determine $I_C \approx 7$ mA for the 15 nm-thick Py waveguide, and calculate the enhancement of the decay length associated with the SOT-induced damping compensation (solid curve in Fig. 4c). The obtained dependence is in a good agreement with the experimental data, demonstrating that the same SOT mechanism enables the generation of spin waves, and simultaneously a significant enhancement of their propagation.

Note that the simulations predict symmetric bidirectional emission of spin waves, despite the well-known non-reciprocity of the Damon-Eshbach spin waves excited in our experiment, and the broken spatial symmetry of the studied structure in the direction normal to the plane. Additionally, the simulated amplitude of the emitted waves amounts to about 10% of the maximum value in the center of the notch (Fig. 5), significantly smaller than 60-70% observed in the experiment (Fig. 3c). We explain these discrepancies by the asymmetric spatial profile of the SOT-driven auto-oscillation mode, whose maximum amplitude is shifted toward one of the edges of the nano-notch, similar to the spatial asymmetry observed in point contact spin torque auto-oscillators[33]. Indeed, the experimental profile of auto-oscillation is clearly shifted in the negative-x (Figs. 3b, 3c) or positive-$x$ (Fig. 3d) direction, depending on the direction of the field. In the simulations, we model this effect by shifting the excitation area by half-width of the nano-notch, resulting in a strongly unidirectional emission, with the asymmetry and coupling efficiency close to those observed in the experiment (Fig. 5).

In conclusion, we have demonstrated a nano-magnonic system that combines all the advantages provided by the spin-orbit torques to locally excite propagating spin waves, and to simultaneously enhance their propagation characteristics. The system is amendable to modifications of structure and geometry, can be implemented with low-damping insulating magnetic materials, and can be easily incorporated as a building block in complex circuits with expanded functionalities. We expect our results to spur significant advances in spin-orbit magnonics, enabling the implementation of efficient spin wave-based computing systems.

**Methods**
**Sample fabrication.** A Pt(4)Py(15)Au(3) film was deposited on the sapphire substrates by high-vacuum sputtering. Here, thicknesses are in nanometers, and the Au(3) capping layer prevented oxidation. A 200-nm wide, 10 nm deep notch was formed in the Py layer by a combination of e-beam lithography and ion milling. The notch was covered with an Au(3) capping layer without breaking the vacuum, to prevent device oxidation. The entire Pt/Py film was then removed by Ar ion milling, except for the 180 nm-wide nanowire protected by the polymer mask defined by e-beam lithography. The nanowire was contacted by Au(60) electrodes. Finally, the Au(3) capping layer was removed by ion milling, and the sample surface was coated with a protective $AlO_x$(10) layer.

**Measurements.** All measurements were performed at room temperature. The avoid Joule heating of the sample, the driving current was applied in 1 μs long pulses with the repetition period of 5 μs. The BLS data were recorded with the temporal resolution of 1 ns, and were averaged over the duration of the current pulse. Micro-focus BLS measurements were performed by focusing light, produced by a continuous-wave single-frequency laser operating at the wavelength of 532 nm, into a diffraction-limited spot. The light scattered from magnetization oscillations was analysed by a six-pass Fabry-Perot interferometer TFP-2HC (JRS Scientific Instruments, Switzerland) to obtain information about the BLS intensity proportional to the intensity of magnetization oscillations at the position of the probing spot. By rastering the spot over the surface of the sample using a closed-loop piezo-scanner, two-dimensional maps of the dynamic magnetization were recorded with the spatial step size of 100 nm. The positioning system was stabilized by custom-designed active feedback, providing long-term spatial stability of better than 50 nm.

**Micromagnetic simulations.** The micromagnetic simulations were performed using the software package MuMax3 (Ref. 34). The computational domain with dimensions of 25 μm ×0.18 μm ×0.015 μm was discretized into 10 nm ×5 nm ×2.5 nm cells. The spin waves were excited by applying a sinusoidal dynamic magnetic field with the amplitude of 1 Oe in the area of the nano-notch. A standard Gilbert damping constant of 0.01 and the exchange stiffness of $1.3\times10^{-11}$ J/m were assumed; the value of the saturation magnetization $4\pi M_0$=9.0 kG was independently determined from the BLS measurements of thermally excited spin waves. The effects of the Oersted field produced by the driving current were incorporated into the spatial distribution of the static magnetic field used in the simulations. The Oersted field was calculated by using COMSOL Multiphysics simulation software (https://www.comsol.com/comsol-multiphysics). Independently measured thickness-corrected resistivities of the Pt and Py films of 275 and 325 nΩ·m, respectively, were used in the calculations.

**Data availability.** The data that support the findings of this study are available from the corresponding author upon reasonable request.

**Acknowledgements:** We acknowledge support from Deutsche Forschungsgemeinschaft, the National Science Foundation of the USA, FASO of Russia (theme "Spin" No. AAAA-A18-118020290104-2), and Russian ministry of Education and Science (project No. 14.Z50.31.0025).

**Author Contributions:** B.D. and V.E.D. performed measurements and data analysis. S.U. and R.F. designed and fabricated the samples, and performed SEM imaging. B.D. and A.B.R. performed micromagnetic simulations. S.O.D. formulated the experimental approach and managed the project. All authors co-wrote the manuscript.

**Additional Information:** The authors have no competing financial interests. Correspondence and requests for materials should be addressed to B.D. (b_divi01@uni-muenster.de).

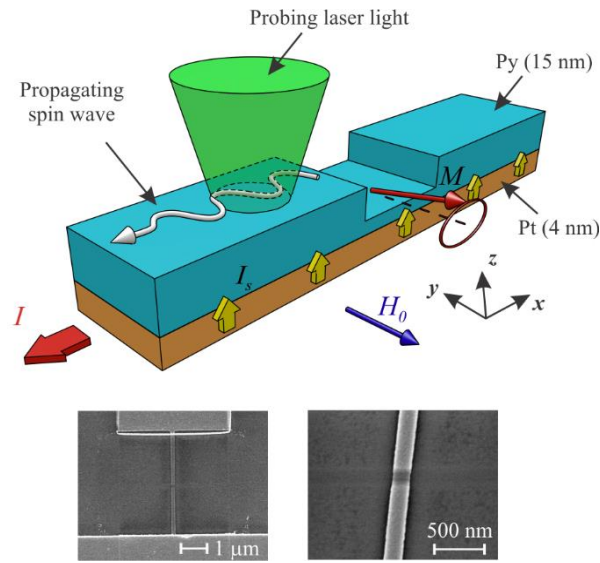

**Figure 1. Schematic of the experiment.** The test devices are 180 nm wide Py(15 nm)/Pt(4 nm) nano-waveguides with a 200 nm wide and 10 nm deep rectangular nano-notch in the center. The injected spin current $I_S$, excites magnetization auto-oscillations in the nano-notch, resulting in the spin wave emission into the waveguide. Insets show the SEM images of the device and of the active nano-notch region.

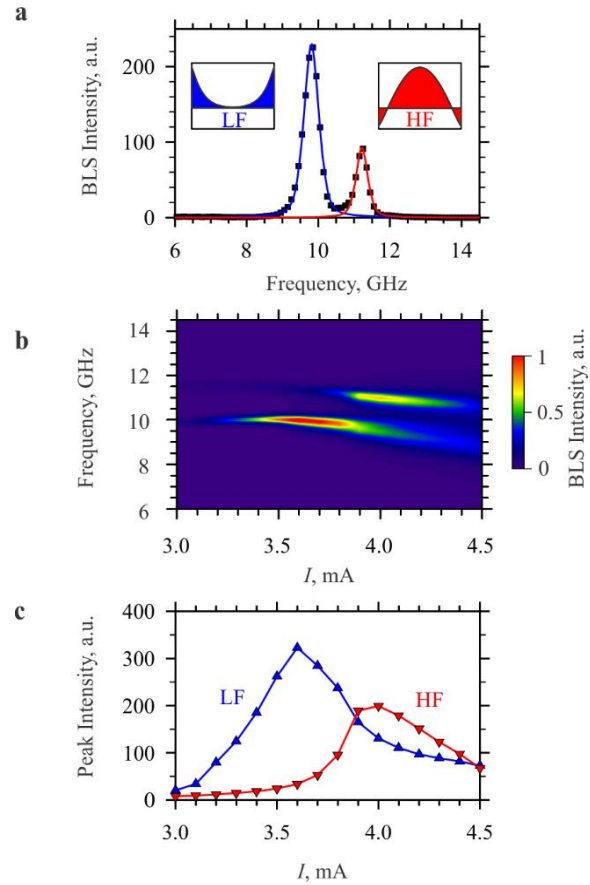

**Figure 2. Characterization of the nano-notch oscillator. a,** Representative BLS spectrum of auto-oscillations measured at *I*=3.8 mA with the probing spot positioned on the nano-notch. Symbols: experimental data, lines: Lorentzian fits of the spectral peaks. Insets schematically show the transverse profiles of the dynamic magnetization corresponding to the low-frequency (LF) and the high-frequency (HF) mode. **b,** Normalized color-coded map of the BLS intensity in the frequency-current coordinates. **c,** Current dependences of the peak intensity for the LF and the HF mode. Symbols: experimental data, lines: guides for the eye. The measurements were performed at $H_0$=2000 Oe.

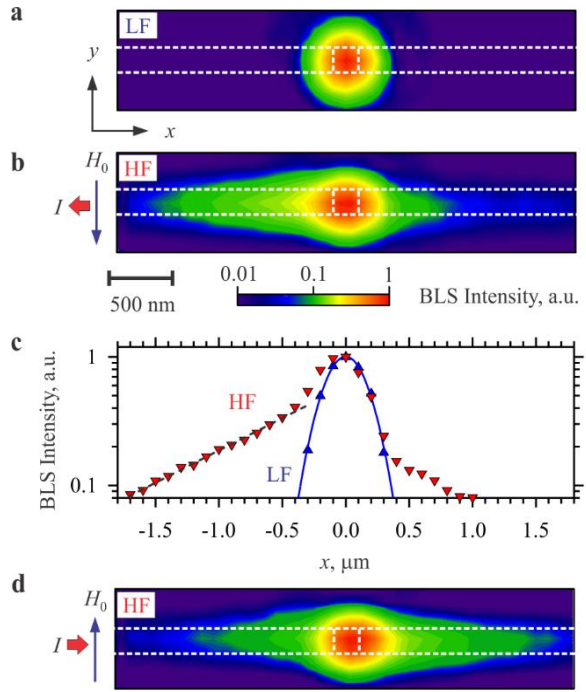

**Figure 3. Spin-wave emission by the nano-notch oscillator. a** and **b**, Color-coded spatial maps of the BLS intensity measured at the frequency of the LF and of the HF mode, as labeled. The maps were recorded at $I$=4.0 mA and $H_0$=2000 Oe. Dashed lines on the maps show the outlines of the waveguide and of the nano-notch. **c**, Symbols: dependence of the BLS intensity for LF and HF modes, as labelled, on the propagation coordinate. Note the logarithmic intensity scale. Solid curve: Gaussian fit of the data for the LF mode, dashed line: exponential fit of the data for the HF mode at $x$<-0.5 μm. **d**, Same as **b**, measured with reversed directions of the static magnetic field and of the driving current.

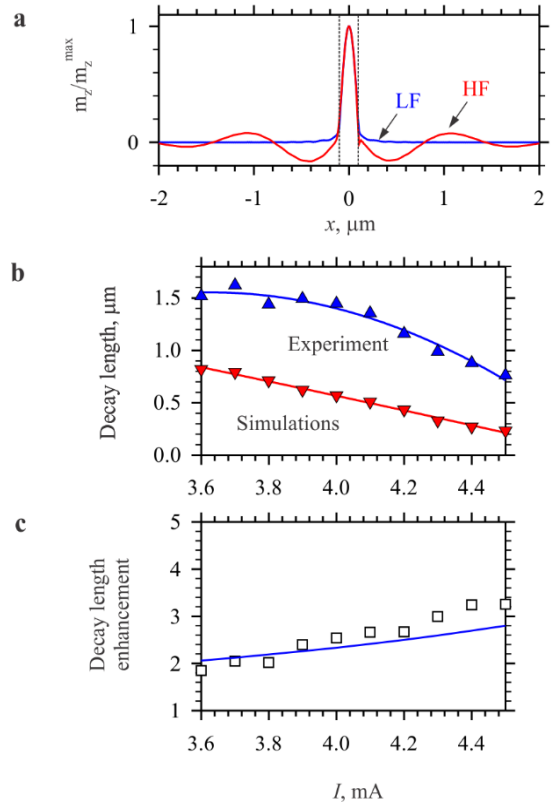

**Figure 4. Micromagnetic simulations of spin-wave emission and propagation. a,** Snapshots of the dynamic magnetization in the waveguide for the excitation at the frequency of the LF (9.97 GHz) and of the HF (11.36 GHz) mode, as labelled. Dashed vertical lines show the edges of the nano-notch. **b,** Current dependence of the decay length of emitted spin waves. Point-up triangles: experimental data. Point-down triangles: results of simulations neglecting the effects of current on the spin wave propagation. Curves: guides for the eye. **c**, Symbols: ratio of the experimental to the simulated value of the decay length. Solid curve: enhancement of the decay length expected from the SOT-induced damping compensation.

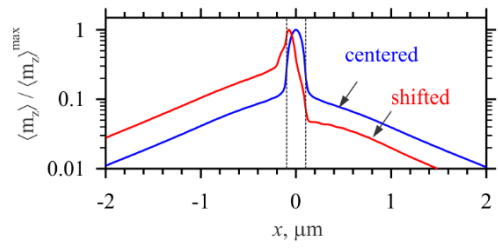

**Figure 5. Emission asymmetry due to the spatial shift of the oscillation region.** Calculated spatial dependences of the time-averaged amplitudes of the dynamic magnetization, for the oscillation region centered on the nano-notch, and for the oscillation region shifted to the edge of the notch, as labelled. Dashed vertical lines indicate the edges of the nano-notch.

# Supplementary information

**Effects of SOT on the effective damping in the 15 nm thick Py waveguide.**

To analyze the effect of SOT on the magnetization damping in the Py waveguide, we utilize BLS spectroscopy of thermal magnetic fluctuations[1]. We position the probing laser spot on the 15-nm thick part of the Py waveguide, at the distance of 1 μm from the nano-notch, while applying a small current $I<3$ mA to induce SOT in the waveguide. At such small currents, spin waves are not emitted by the nano-notch, allowing us to characterize the magnetization dynamics in the waveguide unaffected by the built-in auto-oscillator.

Figure 1a shows the BLS spectra of the magnetic fluctuations, recorded at different currents, as labeled. The shaded spectrum for $I=0$ corresponds to thermal fluctuations, which are always present in the sample at finite temperatures. The magnetic fluctuations become enhanced at finite currents $I>0$, in agreement with the general theory of spin torque[2]. The inverse of the integral fluctuation intensity (Fig. 1b) linearly decreases with $I$, extrapolating to zero at $I_C \approx 7$ mA.

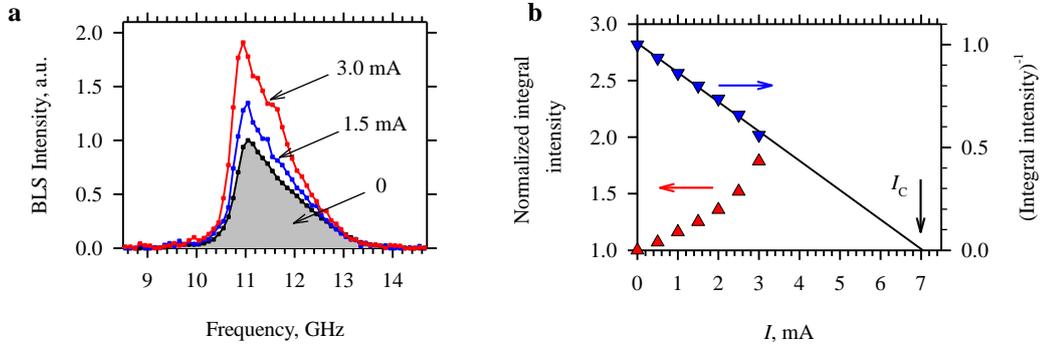

**Figure 1. Effect of SOT on the magnetic fluctuations in the waveguide**. **a**, BLS spectra of the magnetic fluctuations in the waveguide, obtained at the labeled values of current. **b**, Current dependence of the integral intensity of magnetic fluctuations and of its inverse. Symbols: experimental data, line: linear fit of the data for the inverse integral intensity. $I_C$ marks the extrapolated value of current at which the intensity of fluctuations is expected to diverge. The data were obtained at $H_0=2000$ Oe.

According to the spin torque theory, the fluctuation intensity is inversely proportional to the effective current-dependent damping[2]. Therefore, the divergence of the fluctuation intensity at $I=I_C$ is associated with the complete compensation of the magnetic damping by the spin torque. The value $I_C \approx 7$ mA obtained from the fitting is significantly larger than the auto-oscillation onset current for the nano-notch oscillator, consistent with the smaller effects of interfacial SOT on the much thicker Py layer in the extended waveguide region. Based on the linear dependence of the effective damping on current, we estimate that at the operational currents $I=3$-$4.5$ mA of the nano-notch oscillator, the damping in the 15-nm thick waveguide is reduced to 35-55% of its natural value at $I=0$, resulting in the enhancement of the spin wave decay length by up to a factor of three.